\documentclass[aip, amsfonts, amssymb, amsmath , onecolumn, reprint, subscriptadress, showpacs]{revtex4}

\usepackage{amssymb}		 
 \usepackage{amsmath, amsthm, amssymb,amsfonts}
\usepackage[english]{babel}
\usepackage[pdftex]{color, graphicx}
\usepackage{epstopdf} 

\usepackage{natbib}			

\usepackage{caption}
\usepackage{comment}

\usepackage{color}

\graphicspath{{grafici/}}

\begin{document}
\title{ Pressure anisotropy generation \\ in a magnetized plasma configuration with  a shear flow velocity }
\author{\sf {S. De Camillis$^{1,2}$, S.S. Cerri$^{1,3}$, F. Califano$^1$, F. Pegoraro$^1$}}
\affiliation{$~$\\ $^1$ Physics Department ``E. Fermi'', University of Pisa, Largo B. Pontecorvo 3, 56127 Pisa, Italy\\
$^2$ Centre for Plasma Physics, School of Mathematics and Physics, Queen's University Belfast, Belfast BT7 1NN, UK  \\ 
$^3$ Max-Planck-Institut f\"ur Plasmaphysik, Boltzmannstr. 2, D-85748 Garching, Germany }


\begin{abstract}
The nonlinear evolution of the Kelvin Helmholtz instability in a magnetized plasma with a perpendicular flow  close to, or in,  the supermagnetosonic  regime  can produce  a significant parallel-to-perpendicular pressure anisotropy.  This anisotropy, localized inside  the flow shear region, can make the configuration unstable either to the mirror or to the firehose instability and, in general, can affect the development of the KHI.   The interface between the solar wind and the Earth's magnetospheric plasma  at the magnetospheric equatorial flanks provides a relevant setting for the development of this complex nonlinear dynamics. 
\end{abstract}

\maketitle

\section{Introduction} \label{introduction}

The linear and non-linear evolution of the {\em Kelvin-Helmholtz instability} (KHI) plays a central role in the problem of 
the interaction between the solar wind (SW) and Earth's magnetosphere (MS) plasma since a sheared flow configuration naturally arises at the interface of the two plasmas. Multi-spacecraft measurements have provided unambiguous evidence around the equatorial region for rolled-up vortices as well as for the formation of a boundary layer structure arising from the (suggested) nonlinear evolution of the KHI~\cite{Miura1984,Ohtani,Otto2000,Hasegawa2004}. This important mechanism is  thought to be the best candidate for the formation of a mixing layer along the low latitude flanks of the MS (the so-called low latitude boundary layer, LLBL) during SW northwards conditions~\cite{Fairfield,Foullon}, when the SW magnetic field and geomagnetic field are parallel, thus providing an efficient way for the SW plasma to enter the magnetospheric region. Note that a direct reconnection process between field lines advected by the SW and those forming the MS is the dominant mixing process during southwards SW conditions~\cite{Phan}.
On the contrary, reconnection should not take place in the opposite northwards configuration, while experimental data~\cite{Terasawa} suggest  that 
the mixing occurs also during  northward SW condition ~\cite{Fujimoto1998} and that it can be even larger than during southwards SW conditions.
Furthermore, the vortex dynamics is a natural source of secondary magneto-fluid instabilities such as for instance magnetic 
reconnection~\cite{Liu1998,Nakamura2005,Faganello2008c}, KHI  \cite{Smyth2003}, and Rayleigh-Taylor instability (RTI) \cite{Matsumoto2004,Faganello2008a,Tenerani2011}, which are in turn responsible for the evolution of the macroscopic structure. Therefore, the characteristic time scales of these secondary instabilities are crucial for the nonlinear evolution of the primary KHI, as they compete with the  hydrodynamic process  of vortex pairing by driving disruptive processes inside and outside the vortices and thus leading to the formation of a mixing layer. 

Previous   articles  have analyzed the conditions for the onset and  the characteristic time-scales of each instability emphasizing their influence on the evolution of the mixing layer during the SW-MS interaction (see~\cite{henri} and references therein). All the simulations presented in these articles have focused their attention on the magneto-fluid dynamics of a sheared velocity  {plasma} configuration assuming, for the sake of simplicity, an isotropic pressure tensor and either an adiabatic or an isothermal relation between the scalar pressure and the density. Nevertheless, space plasmas are often characterized by the presence of a background magnetic field which naturally introduces an asymmetry in the plasma dynamics along and perpendicular to the field. Thus, in the presence of a relatively strong magnetic field, which is the case for the LLBL introduced above, an isotropic pressure tensor, as  typically used in a fluid framework, is not appropriate and one should revert  to a kinetic description. However, even with the most powerful super-computers of the latest generation, a fully kinetic approach to solving the Vlasov equation remains very difficult and computationally demanding  and probably not the best solution when the system geometry is quite complex and the dynamics remain at relatively low frequencies.

An alternative approach with respect to a kinetic description is to develop fluid models that extend the large-scale magnetohydrodynamic (MHD) approach towards regimes where the effects related to the presence of fluctuations at scale lengths comparable to the ion Larmor radius 
$\rho_{_i}$  (or to the ion skin depth $d_i$) are included (see e.g., ~\cite{Rosenbluth1962,Roberts1962} and, more recently, see e.g.,~\cite{Snyder1997}). These models can be used provided the characteristic frequencies of the plasma remain``smaller" than the ion cyclotron frequency and phenomena related to the ion cyclotron resonance remain negligible. Extended models have a wide range of applications, from laboratory \cite{kunz2015} and space plasma turbulence~\cite{Passot2007} to magnetic reconnection studies. More generally, they can describe plasma configurations where the energy injected at large scales cascades efficiently towards the ion Larmor radius and/or the ion inertial scale. This {is  the case of  the} problem of the formation of the LLBL between the SW and the MS because of the efficiency of {the} secondary instabilities and {of the} non-linear interactions in generating smaller and smaller scales up to a fully turbulent state~\cite{Karimabadi2013}, but also because of the role of the pressure tensor in the presence of a shear layer~\cite{Cerri_PressTens}  as  discussed in the present  article.

Here we will adopt an extended fluid model valid in the small Larmor radius approximation, $k \rho_{_i} \ll 1$ recently presented in Ref.~\cite{ext_two-fluid_model} allowing for the inclusion of two distinct scalar pressures, $p_\|$ and $p_\perp$, one along the magnetic field and the other perpendicular to it, respectively, as well as for the inclusion at first order of the so-called Finite Larmor Radius (FLR) effects. This approach allow us to include a very important feature  in {the plasma dynamics  i.e.,} the development of pressure anisotropy { driven  in this case  by the non-linear evolution} of KH vortices. 

	Aside  from   the importance {\em per se} of including a non-isotropic pressure in the model, { the} development of pressure 
anisotropy could {drive}  the plasma  towards  less stable configurations. One of the most important pressure anisotropy instabilities that could arise in such conditions is the Firehose instability (FHI) {that can make} transversal perturbations propagating along the magnetic field grow exponentially in time. 
The FHI {onset} condition is met when the pressure anisotropy exceeds (twice) the magnetic field tension. 
 In order to  calculate the FHI growth rate  correctly  we would need to  consider a 3D configuration beside including the  FLR effects in the model equations. Although the results  presented in this article are restricted to a 2D configuration,  in view of a future comparison, we include  the FLR terms  in our model and  do not  simply consider   a parallel versus perpendicular pressure which could be achieved by means of a relatively simpler Chew-Goldberger-Low (CGL)-like approach \cite{chew}. 
 In the context of the SW-MS interaction, the FHI {can}  play a central role during the development of vortex structures since it {can}
contribute as a secondary instability to those disruptive processes developing within the K-H vortices. Indeed, if the onset of the FHI 
is reached during the early nonlinear stage of the primary KHI, it { would compete} with the characteristic hydrodynamic 
vortex time-scale becoming one of the dominant process in the subsequent nonlinear evolution. In particular{, by driving 
the growth of transverse magnetic perturbations,  this instability could  generate} a locally strong in-plane magnetic field component {eventually   radically
 modifying  the} plasma dynamics.

{In addition to the onset of the FHI, we  also  find  that the plasma can  develop  conditions that lead in principle to} the onset of the mirror instability (hereafter MI). However, the quantitative calculation of the MI development cannot be made {within} a fluid approach, even extended to FRL effects, as it needs a fully kinetic treatment in order to account for its stabilization at small wave lengths {as well as to  calculate the growth rates correctly.}

The paper is organized as follows: in Sec.\ref{sec_equations} the governing equations adopted in our plasma  model are described while in  Sec.\ref{InCon} 
the simulation initial conditions  are given and in Sec.\ref{sec_parameters}  the plasma parameters.   In Sec.\ref{sec_equations} the result of the numerical integration of our plasma model equations are presented and discussed in the case where a small in-plane   magnetic field component is initially present ( Sec.\ref{magn_reg}) and in the case of a purely perpendicular ``guide'' magnetic field ( Sec.\ref{gf_reg}). Finally the conclusions are given in  Sec.\ref{conclusions}.

\section{  {Model} equations} \label{sec_equations}

We adopt the so-called ``{\em e}TF model'' presented in Ref. \cite{ext_two-fluid_model} in which the plasma is assumed { to be made of }   two distinct, collisionless fluids, ions and electrons ({indices  $e$ and $i$}, respectively) interacting through the electromagnetic fields. In 
{this  model we} assume quasi neutrality, $n_e \simeq n_i = n$, and massless electrons. {We do not} adopt a standard isotropic pressure closure (e.g. an isothermal or adiabatic law) {and}  consider instead the case of a 
gyrotropic pressure for both species plus ions first-order finite Larmor radius (FLR) agyrotropic corrections. The gyrotropic components of the pressure tensor
are evolved in time (see Appendix  A in Ref. \cite{ext_two-fluid_model} for details) while the FLR terms are determined according to Ref.\cite{ext_two-fluid_model}. 

We normalize all quantities using as references  the  values  of the   ion quantities at the simulation box boundary  where the plasma is locally homogeneous:  the ion inertial scale-length $d_i = c / \omega_{pi}$ and 
the ion gyro-frequency $\Omega_{ci}^{-1}$ and, consequently,  the Alfv\'en velocity ${\bar c}_{_A} = B/\sqrt{4 \pi n_i m_i}$. The parallel and 
perpendicular pressure components  are normalized to the total reference pressure ${\bar p}_{_0} = {\bar p}_{\alpha\|} + {\bar p}_{\alpha \perp}$. Here parallel and perpendicular refer to the direction of the  direction  of the magnetic field.

 The model equations read:

 \begin{equation}\label{eq:cont_de0}
  \frac{\partial n}{\partial t}\ +\ \mathbf{\nabla} \cdot\left(n \mathbf{U}\right) \ =\ 0
 \end{equation}
 \begin{equation}\label{eq:motion_de0}
  \frac{\partial\left(n\mathbf{U}\right)}{\partial t}\ +\ 
  \nabla\left[n{\bf UU} + \boldsymbol{\Pi}_{\rm B} +
  \boldsymbol{\Pi}_e^{(0)} + \boldsymbol{\Pi}_i^{(0)} + \boldsymbol{\Pi}_i^{(1)}\right]\ =\ 0
 \end{equation}
 \begin{equation}\label{eq:Ohm_de0}
  {\bf E}\ =\ -{\bf{U}}\times{\bf B}\ +\ \frac{{\bf J}\times{\bf B}}{n}\ -\ \frac{\nabla\boldsymbol{\Pi}_e^{(0)}}{n}
 \end{equation}
 \begin{equation}\label{eq:Maxw_de0}
  \frac{\partial\mathbf{B}}{\partial t}\ =\ -\ \mathbf{\nabla}\times\mathbf{E}\,  ; \;\;\;
  \mathbf{\nabla}\times\mathbf{B} =  {\bf J}
 \end{equation}
\begin{equation}\label{eq:pperp_i_de0}
 \frac{\partial p_{i\perp}}{\partial t}\ +\ \nabla\left(p_{i\perp}{\bf U}\right)\ =\
 -p_{i\perp}\boldsymbol{\tau}\,{\bf:\nabla U}\ -\ \boldsymbol{\Pi}_i^{(1)}{\bf:\nabla U}
\end{equation}
\begin{equation}\label{eq:pperp_e_de0}
 \frac{\partial p_{e\perp}}{\partial t}\ +\ \nabla\left[p_{e\perp}\left({\bf U}-\frac{{\bf J}}{n}\right)\right]\ =\
 -p_{e\perp}\left[\boldsymbol{\tau}\,{\bf:\nabla}\left({\bf U}-\frac{{\bf J}}{n}\right)\right]
\end{equation}
\begin{equation}\label{eq:ppar_i_de0}
 \frac{\partial p_{i\|}}{\partial t}\ +\ \nabla\left(p_{i\|}{\bf U}\right)\ =\
 -2p_{i\|}\, {\bf bb:\nabla U}
\end{equation}
\begin{equation}\label{eq:ppar_e_de0}
 \frac{\partial p_{e\|}}{\partial t}\ +\ \nabla\left[p_{e\|}\left({\bf U}-\frac{{\bf J}}{n}\right)\right]\ =\
 -2p_{e\|}\left[{\bf bb:\nabla}\left({\bf U}-\frac{{\bf J}}{n}\right)\right]\ ,
\end{equation}
where $p_{\alpha\|}$ and $p_{\alpha \perp}$ are the pressure components of the $\alpha$ species, parallel and perpendicular to 
the magnetic field, respectively, $\mathbf{b}=\mathbf{B}/B$ is the unit vector along the local magnetic field, $\boldsymbol{\tau}\equiv{\bf I-bb}$ is the projector in the perpendicular plane and $\boldsymbol{\Pi}_{\rm B}=(B^2/2)\left(\boldsymbol{\tau}-{\bf bb}\right)$, \,$\mathbf{\Pi}_\alpha^{(0)}=p_{\alpha\perp}\big({\bf I -bb} \big) + p_{\alpha\|}{\bf bb}$ are the magnetic pressure tensor and the gyrotropic part of the pressure tensor of the $\alpha$ species, respectively. Finally $\boldsymbol{\Pi}_i^{(1)}$ is the first order ion gyroviscosity tensor with components $\Pi_{i,lm}^{(1)}$ as given in Ref.\cite{ext_two-fluid_model}. 

In the following we present 2D simulations of the KHI in the limit of a relatively strong magnetic guide field with 
$\mathbf{B} \simeq B \mathbf{e}_z$. The set of equations (\ref{eq:cont_de0})-(\ref{eq:ppar_e_de0}) is integrated 
numerically in a 2D $(x,y)$ domain with $L_x = 420$ and $L_y = 60 \pi$ using $N_x = 2048$, $N_y = 1024$ corresponding 
to $dx \simeq dy \simeq 0.2$. The simulation domain is periodic along 
the $y$-direction and open in the $x$-direction where we impose transparent boundary conditions in order 
to let sonic and Alfv\'enic perturbations, generated by the dynamics of the KHI far from the simulation $x$-boundaries,  {leave}  the simulation box. 
In order to apply the open transparent boundary conditions scheme (see Ref. \cite{faganello:being} for details), we force 
the system to relax towards the $x$-boundary to an isotropic pressure condition, $p_{\alpha\perp} = p_{\alpha\|}$, as set at the initial time. 
This is achieved by smoothing the pressures of each species  ($\alpha=i,e$) to the boundary value 
$p_\alpha = 1/3 [2 p_{\alpha\perp} + p_{\alpha\|}]$ using a hyperbolic tangent profile. Similarly, 
the velocity field is gradually smoothed in the $x$-boundary region, so that the electron velocity becomes equal to 
the ion velocity. Finally, the stability of the code is controlled through numerical filters which suppress 
{very short wavelength fluctuations}. 
\bigskip

\section{Initial conditions}\label{InCon} 

{We consider  an initial plasma configuration with a sheared velocity field unstable} against the KH instability. For numerical convenience, 
we make a Galilean transformation {to a frame moving  with  velocity  $U_0/2$, so} that the flow velocity is symmetric with respect to the shear center 
$x=0$, see Eq.(\ref{initial_velocity}). {Since} the initial density is symmetric with respect to $x=0$ (and varies {by} no more 
than $5\%$ so that it can be assumed to be  nearly constant), the vortices generated by the KHI will be at rest in the {simulation}  frame.
The overall velocity jump $U_0$ is given by {the profile}
\begin{equation}
{\bf U}_{in} = \frac{U_0}{2} \tanh \left( \frac{x - x_0}{L_{eq}} \right) \; {\bf e}_y
\label{initial_velocity}
\end{equation}
where $x_0=L_x/2$ and $L_{eq}$ is the characteristic scale-length of variation of the equilibrium. In the {simulations there }
are about two order of magnitude between $L_x$ and $L_{eq}$, meaning that the boundary is very far from the region where 
the dynamics develops. The initial magnetic field reads
\begin{equation}
{\bf B}_{in} = B_0(x) \sin\theta \; {\bf e}_y + B_0(x) \cos\theta \; {\bf e}_z
\label{initial_magneticfield}
\end{equation}
where $\theta$ is the angle of the magnetic field vector with respect to the z-axis. The angle $\theta$ is introduced in order to allow 
for an in-plane component of the magnetic field, as is of interest for the investigation of the  magnetosphere solar wind interface at the Earth's  flanks.
It is assumed to be small enough 
so as { not to }inhibit the linear development of the KHI as well as to allow for the formation of the vortex chain in 
the non-linear phase \cite{faganello:being}.
We take an initial equilibrium condition where the magnetic field profile, $B_0(x)$, the parallel and perpendicular pressure profiles, $ p_{\alpha,\|}(x)$ and $p_{\alpha,\perp}(x)$, ensure that the total pressure is constant, $\nabla\cdot(\boldsymbol{\Pi}_e^{(0)} + \boldsymbol{\Pi}_i^{(0)} + \boldsymbol{\Pi}_i^{(1)} + \boldsymbol{\Pi}_{\rm B}) = 0$, with no magnetic tension, $({\bf B} \cdot \nabla) {\bf B} = 0$ (the details on the equilibrium profiles can be found in Ref.\cite{ext_two-fluid_model}).  

{At $t=0$ we introduce an initial random noise of small amplitude ($\sim 10^{-4}$) given by the superposition of compressible and incompressible modes with  a spectrum of wave vectors $k_m$  along the flow direction and centered  in the region of the velocity shear.}

\section{Plasma simulation  parameters}\label{sec_parameters}

In a magnetized plasma a sheared velocity configuration develops   fluctuations propagating at the  fast magnetosonic speed, given by the square root  of the square of the  Alfv\'en velocity, $v_{_A}^2=B^2/4\pi nm_i$ and  of the sound speed in the plane perpendicular to the magnetic field, $c_{s \perp}^2 = \gamma_{_{\perp}} (T_{i \perp} + T_{e \perp})/m_i$,  where $\gamma_{_{\perp}}$ is the polytropic index perpendicular to ${\bf B}$. In an anisotropic plasma  we distinguish two different polytropic indexes 
$\gamma_{_{\perp}}$ and $\gamma_{_{\parallel}}$, depending on the direction with respect to the ambient magnetic field. 

As long as the in-plane magnetic components are negligible as compared to the perpendicular one, $B_z$, the pressure equations (\ref{eq:pperp_i_de0})-(\ref{eq:ppar_e_de0}) are equivalent to polytropic closures with polytropic indexes for each species, $\gamma_{_{\parallel}} = 1$   along $z$ and $\gamma_{_{\perp}} = 2$ in the $x$-$y$ plane \cite{Cerri_PressTens}.  
The possibility to distinguish between  $\gamma_{_{\parallel}}$ and $\gamma_{_{\perp}}$ is important in 
the presence of a shear flow since the plasma compressibility affects  the development of the KHI even during  its
initial linear phase (in general, the greater the  compressibility the smaller the KH growth rate). Moreover, in the non linear regime 
the plasma compressibility determines the  { plasma regime} where different processes  such as vortex pairing, secondary fluid-like instabilities, 
magnetic reconnection compete~\cite{henri}.

The critical parameter that distinguishes the different regimes of the plasma dynamics  is 
the {\em in-plane fast magnetosonic Mach number} given by the ratio
\begin{equation}  \label{magnetofast_mach_number}
M_{f \perp} = U_0 / \sqrt{v_A^2 + c_{s \perp}^2}
\end{equation}
In the present  {article}  we investigate the  pressure anisotropy generation passing from the {\em subsonic} to 
the {\em supermagnetosonic } regime, i.e., for $M_{f \perp}$ less or greater than unity. We recall that, { after 
the KHI has  generated a vortex chain}, the rolled-up vortices act as an "obstacle" with respect to the initial 
mean flow  and can generate shocks: see Ref.~\cite{palermo2} on 
the shock formation by a vortex structure. 

We  use  the following plasma parameters, 
\begin{equation}
B_0 = 1 ; \;\;\; n_0 = 1 ; \;\;\;  L_{eq} = 3 ; \;\;\; \theta = 0.02   
\end{equation}
where the quantities $B_0$ and $n_0$ are the reference   values of the magnetic field and number density  
far away from the shear layer. \
In the central region, because of the FLR equilibrium condition\footnote{We remind  that the central density variation is a characteristic feature of 
the FLR equilibrium in contrast to the standard MHD approach where the density would be strictly constant. }
the density varies only by a few percent with respect to its reference  value (which is the same on both sides). 
Such a density variation is very small and thus  {not sufficient}  to cause, after the development of the primary KH vortices, 
the onset of Rayleigh-Taylor like secondary instabilities (driven by the centrifugal acceleration in the side the rotating vortices)  or 
to influence the non linear dynamics of the plasma with respect to what is known in the MHD limit in the case of a constant density shear flow configuration.

\section{Numerical results}\label{sec_results}
\subsection{Magnetized regime}\label{magn_reg}

We present a set of numerical simulations that show  that the KHI  
can drive the formation of plasma pressure anisotropy   when the supermagnetosonic  regime is approached. This anisotropy can, in its turn, become  a source for 
the development of secondary instabilities, first of all the firehose and the mirror instabilities. 
We consider three values of the flow velocity $U_0 = 2, \, 4$ and $6$ (in dimensionless units) as summarized in Table  \ref{tab:mach_uniform}. These values correspond to 
the subsonic, { the intermediate and  the} supermagnetosonic  regime, $M_{f \perp} = 0.58$,  \, $1.1$, and  $1.7$, respectively. For each of these regimes, we consider the case 
with  or without  a small in-plane component of the magnetic field aligned with the flow, $\theta = 0$ or $\theta = 0.02$, respectively. \,
The  simulations have been performed by integrating the model equations presented in Section \ref{sec_equations}.  In this context we note that  the expressions for the first-order FLR corrections implemented in 
the code are valid as long as  the in-plane component of the magnetic field 
is small compared to the guide field, i.e. in the limit of $B_\perp\ll B_z$.  This condition represents a constraint  
to be taken into account during the full dynamics, since the plasma differential compression and stretching  may 
eventually lead to a local amplification of the in-plane magnetic field $B_\perp$ up to relatively large values for which the validity of 
the model could be called in question. This point is discussed at the end of this Section. The parameters of the different simulation runs are listed in Table \ref{tab:mach_uniform}.

\begin{table} [!h] 
\begin{tabular}{|c|c|c|c|}
\hline 
\hspace{1.5cm} & \hspace{0.1cm} $U_0$ \hspace{0.1cm} & \hspace{0.1cm} $\theta$ \hspace{0.1cm} & \hspace{0.1cm} $M_{f \perp}$ \hspace{0.1cm} \tabularnewline
\hline 
Run 1  & 2 & 0 & 0.58  \tabularnewline
\hline 
Run 2  & 2 & 0.02 & 0.58  \tabularnewline
\hline 
Run 3  & 4 & 0 & 1.1  \tabularnewline
\hline 
Run 4  & 4 & 0.02 & 1.1  \tabularnewline
\hline
Run 5  & 6 & 0 & 1.7  \tabularnewline
\hline 
Run 6  & 6 & 0.02 & 1.7  \tabularnewline
\hline
\end{tabular}\caption{Main simulation parameters}
 \label{tab:mach_uniform}
\end{table}


{In order to investigate the  process  of anisotropy generation by a shear flow quantitatively},  we  refer to the pressure anisotropy of each species  defined as  the difference between the out-plane and in-plane 
(parallel and perpendicular) component of the pressure tensor,
\begin{equation}\label{panis}
\Delta P_\alpha = p_{\alpha\|} - p_{\alpha\perp} \quad\ (\alpha=i,e)
\end{equation}
normalized to 
the reference total pressure. We also  define the ``$\beta$-anisotropy" as
\begin{equation}\label{anisfield}
\Delta \beta_{anis} = 2 \left( \Delta P_i + \Delta P_e \right) / B^2 \;\; 
\end{equation}
 which is the   parameter  that controls  the onset of the FHI and MI. It 
depends both on { the pressure anisotropy and on the } magnetic pressure. 
\,
{The reason  why plasma compression/rarefaction plays} a key role in the anisotropy formation can be 
seen  simply from the fact  that, due  the two different polytropic indices in the pressure equations: $\gamma_{_{\parallel}}  = 1$ and $\gamma_{_{\perp}}  = 2$,  any density variation  
will lead to  different  in-plane and out-plane  variations of  an initially isotropic  pressure and thus to the formation of pressure anisotropy and of  a $\beta$-anisotropy. The latter  may be further amplified   by a reduction of the magnetic pressure.  

At $t=0$ we include  a small in-plane component of the magnetic field, Run 2, 4 and 6. 
This is a typical configuration studied extensively in the  literature, in particular in the subsonic regime, because of its 
relevance to the problem of the dynamics at the low latitude magnetosphere boundary layer (see for example 
Ref.~\cite{henri} and references therein). 

\begin{figure} [!t]
\includegraphics[width=10cm]{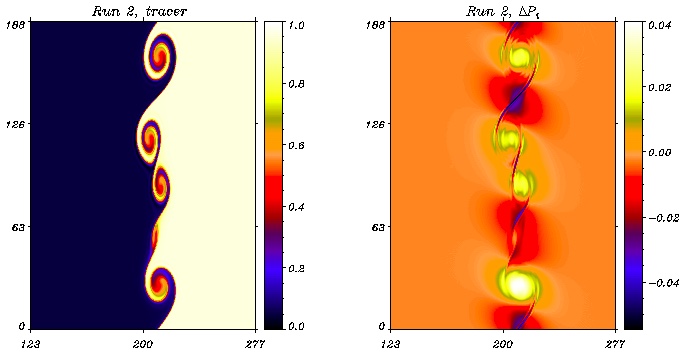}
 \caption{Subsonic regime (Run 2). The tracer and the ion pressure anisotropy, left and right frame respectively, at $t=375$.}
 \label{fig1}
 \end{figure}

In Fig. \ref{fig1}, left frame, we show {the shaded contours of 
the passive tracer representing the two   the solar wind  (yellow) and the magnetosphere (blue)  plasmas for the subsonic regime, run 2, at $t=375$}. 
The tracer field highlights the plasma dynamics emerging from the development of the KHI. We observe 
the formation at the end of the linear phase  of four main vortices corresponding to the Fast Growing Mode, 
m=4, of our configuration. Two of them have already started to pair. In the right frame we {show}, at 
the same time, the ion pressure anisotropy $\Delta P_i$ as defined in Eq.(\ref{panis}). 
\begin{figure} [h!]
\includegraphics[width=10cm]{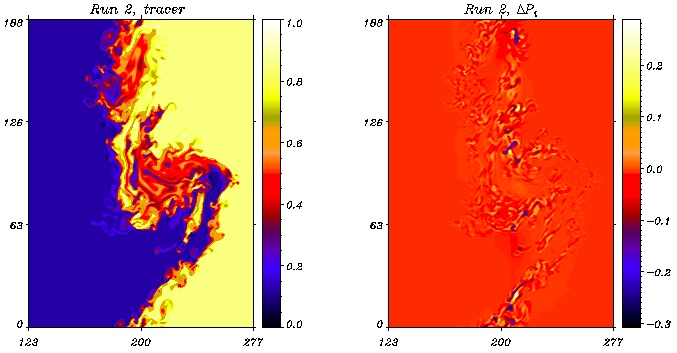}
 \caption{Subsonic regime (Run 2). The tracer and the ion pressure anisotropy, left and right frame respectively, at $t=670$}
 \label{fig2}
 \end{figure}
\begin{figure} [t!]
\includegraphics[width=10cm]{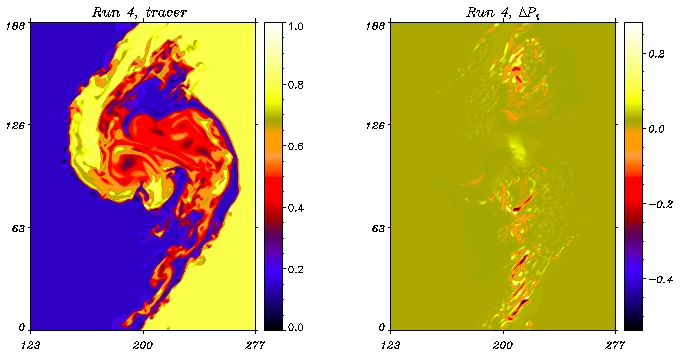}
 \caption{Intermediate regime (Run 4). The tracer and the ion pressure anisotropy, left and right frame respectively, at $t=540$}
 \label{fig3}
 \end{figure}
{The}  pressure anisotropy {grows in {absolute value}  in correspondence to the compressed or rarefied regions. 
Plasma rarefaction occurs mainly inside the vortices during their formation, see Refs. \cite{miura:compressible,palermo}, while 
compression is mainly observed around the vortices in the form of a ribbon structure (surrounded by thin strips of 
rarefied plasma) or in between the vortices. In the same regions the 
magnetic field is also strongly { reduced because of the plasma expansion }  or { enhanced because of plasma compression}  thus affecting  the value of  
$\Delta \beta_{anis}$, see  Eq.~(\ref{anisfield}). Depending on the nature of the density variation, rarefaction or 
compression, the sign of the pressure anisotropy, and  thus  of $\Delta \beta_{anis}$, will be either positive or negative thus 
selecting the possible further secondary instability, a firehose or  a mirror instability, respectively: in the case of a rarefaction, 
$p_{\alpha\perp}$ becomes smaller than $p_{\alpha\|}$, corresponding to a positive  $\Delta \beta$, 
see  Eq.~(\ref{anisfield}) while  n the case of a compression, $p_{\alpha\perp}$   becomes
larger than $p_{\alpha\|}$, corresponding to a negative  $\Delta \beta_{anis}$.}

In Fig. \ref{fig2} we {show}  the same quantities as in Fig. \ref{fig1} at { the  later time  $t=670$,  when 
the pairing mechanism has  generated } one large big vortex corresponding to the largest size allowed by 
the numerical domain. The final vortex is characterized by a boundary profile dominated by small scale structures and filaments, 
see left frame, resulting from the development of local secondary instabilities. In the right frame, we see that the ion pressure 
anisotropy peaks inside {several}  ion-scale structures along the boundary. These structures are relatively {stable, are advected by 
the flow and maintain} their own anisotropy. Except for these   very localized structures, 
the ion pressure anisotropy is approximately negligible elsewhere. 

The capability of the system {to generate} plasma anisotropy by rarefaction or { compression} 
{ increases }  at larger flow velocities, for instance  {at} $U_{_0}=4$ and $U_{_0}=6$ (see Table  \ref{tab:mach_uniform}). In 
Fig. \ref{fig3} and in Fig. \ref{fig4} (left and middle frame) we show  {the same quantities as in the previous figures during the non-linear regime long after saturation
for Run 4 and Run 6. These runs correspond to an intermediate and to a supermagnetosonic  regime, respectively}. As discussed in Ref. \cite{palermo}, at such values of 
the mean flow velocity $U_{_0}$ there is a transition towards the supermagnetosonic  regime. 
The formation of shock structures is the main signature of this transition. Shocks cannot be observed in the tracer  distribution but  are clearly visible in the ion pressure anisotropy (see Fig. \ref{fig4} middle frame) as well as in the other plasma quantities (not shown here)  both on the left and on the right side of the vortices.  Aside for the shock formation, the main difference with respect to the subsonic regime is  that the development of secondary instabilities becomes increasingly ineffective 
and that  extended regions are formed where the plasma is {either rarefied or compressed 
(both in terms of the fluid and of the magnetic pressure) in particular inside the vortices or at their boundaries. Most of the plasma anisotropy is generated inside these regions.

In Fig. \ref{fig4}, right frame, we show 
the shaded iso-contours of the ratio between the in-plane and the out-plane magnetic field component for the supermagnetosonic  regime, Run 6.
We see that significant local  increases  of the in-plane magnetic field are produced inside the regions where the plasma is stretched and compressed while in 
the vortex central region, where we observe the formation of ion (and electron) pressure anisotropy (see, e.g., 
same figure, middle frame), the in-plane magnetic field remains very small. This is particularly relevant concerning 
the validity of the ``eTF model'' applied to the  process of anisotropy generation since the model is obtained in the limit of a magnetic 
field perpendicular to the plane of the dynamics. This point will be further discussed in the following.   

\begin{figure} [t!]
\includegraphics[width=4.9cm]{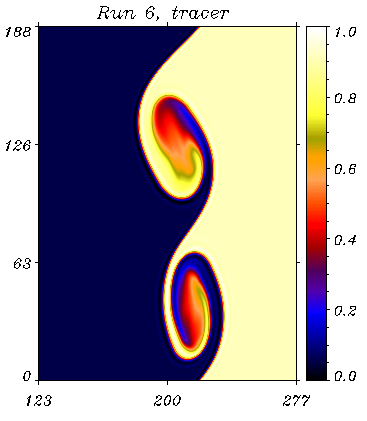}
\includegraphics[width=5cm]{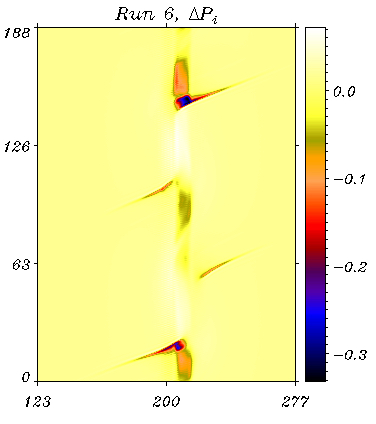}
\includegraphics[width=5cm]{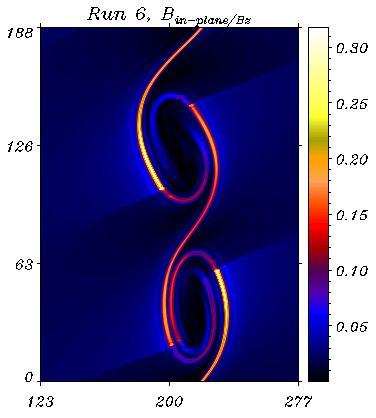}
\caption{supermagnetosonic  regime (Run 6). The tracer, the ion pressure anisotropy and the ratio between the in-plane and out-plane magnetic field, left, central and right frame respectively, at $t=374$}
 \label{fig4}
 \end{figure}
\begin{figure} [t!]
\includegraphics[width=4.5cm]{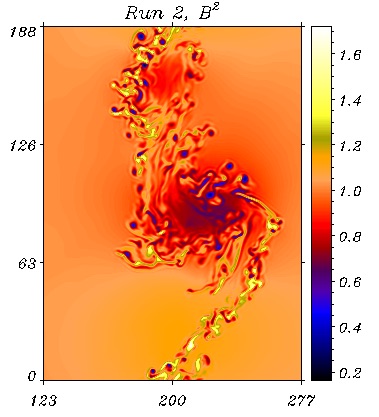}
\includegraphics[width=4.5cm]{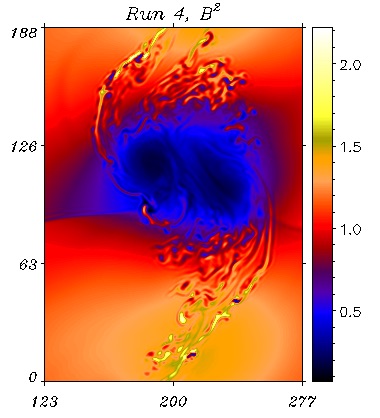}
\includegraphics[width=4.5cm]{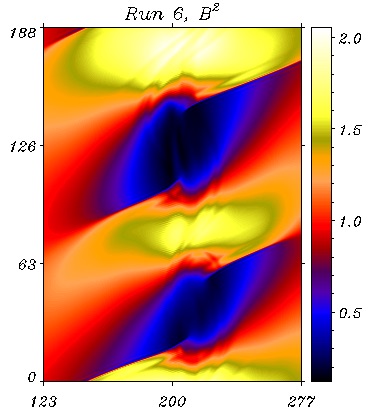}
 \caption{The magnetic pressure for the subsonic ($t=660$), intermediate ($t=520$), supermagnetosonic  ($t=374$) regime, Run 2, 4, 6, left, middle and right frame, respectively}
 \label{fig5}
 \end{figure}

In order to stress   the importance of the {expansion/compression} process associated with the vortex formation, { in Fig. \ref{fig5} we show
the square of the  magnetic field amplitude} $B^2$ for the subsonic, { the intermediate and  the supermagnetosonic  regimes}, Run 2, 4 and 6, respectively, at 
the end of the saturated phase. The magnetic pressure ($\propto B^2$) is a direct signature of the plasma compression.
By comparing these two pictures with the left frame of  Fig. \ref{fig2}, \ref{fig3} and \ref{fig4}, we see, as already discussed, that 
the vortices correspond to 
the regions of plasma rarefaction while the plasma is compressed in between. Quantitatively, the maximum (minimum) vortex 
rarefaction/compression values are comparable. {This expansion/compression effect  increases as 
we move towards the supermagnetosonic  regime as the extension of the expanded/ compressed region increases}. 
In the subsonic regime where secondary instabilities develop very efficiently, 
the peak values of plasma {expansion/ compression} are reached mainly inside small scale  structures or at 
the vortex edges, (see Fig. \ref{fig5}, left frame). On the { contrary in the intermediate and supermagnetosonic  regimes 
the plasma becomes strongly rarefied all over  the vortex region and}  compressed in between the vortices, 
as shown in Fig. \ref{fig5}, middle and right frame, respectively.

In Fig. \ref{fig6} we show the shaded iso-contours of $\Delta \beta_{anis}$ as defined in Eq.(\ref{anisfield}) for 
the three regimes, Run 2, 4 and 6, respectively. We see that the dynamics naturally lead to configurations characterized by 
the presence { in the central region of relatively strong $\beta$-anisotropy values  that increase} as the flow intensity increases. 
Both positive and negative {values of $\Delta \beta_{anis}$ are produced which, in principle, could drive the plasma towards 
the development of the firehose or  of the mirror instability}. However in the subsonic regime the plasma anisotropy is  limited  to small  regions  so that a transition to a $\Delta \beta_{anis}$ unstable dynamics  is unlikely. On the other hand  in the intermediate and supermagnetosonic  regimes    $\Delta \beta_{anis}$  can become large  over relatively extended regions so that a transition towards a firehose or mirror unstable dynamics is expected in these regimes. 

\begin{figure} [t!]
\includegraphics[width=5.125cm]{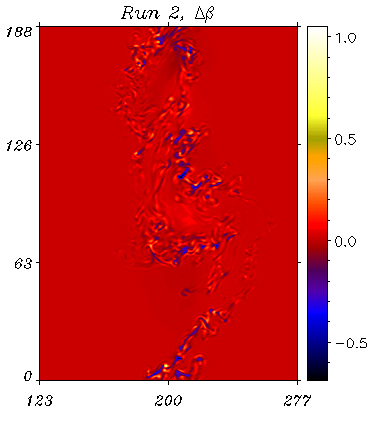}
\includegraphics[width=5cm]{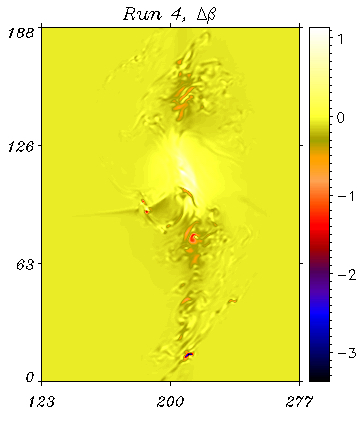}
\includegraphics[width=5cm]{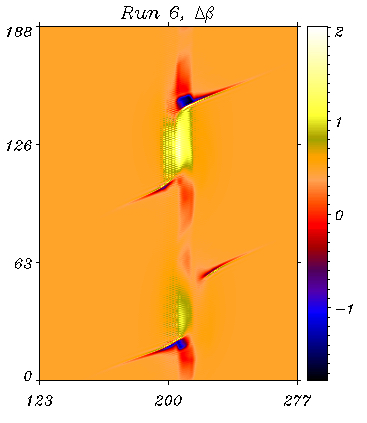}
 \caption{The plasma  anisotropy as defined in Eq. (\ref{anisfield}) for the subsonic ($t=660$), intermediate ($t=520$), supermagnetosonic  ($t=374$) regime, Run 2, 4, 6, left, middle and right frame, respectively}
 \label{fig6}
 \end{figure}

We see that these simulation results indicate that the  nonlinear evolution of the KHI   in the intermediate and  supermagnetosonic  regime can 
produce positive anisotropy values close to  the firehose threshold and that, in addition,  this anisotropy  continues to increase with time. However, 
the 2D  configuration  assumed { in the present investigation   does not permit a correct study of  the  transition to} a firehose unstable regime and 
thus it would be incorrect to continue the analysis of the plasma  evolution  after this threshold is reached. A similar 
argument holds for the negative anisotropy values where it would be  incorrect even to refer to an  instability 
threshold since the mirror instability must be addressed by using a kinetic model.

\subsection{Exact Guide field regime}\label{gf_reg}

In   Sec.\ref{magn_reg} we have studied the process of  plasma anisotropy generation in the presence of an initial 
magnetic field  at a small  angle $\theta$ as defined in Eq.(\ref{initial_magneticfield}). This corresponds to a guide magnetic 
field directed {along $z$} with a small in-plane component parallel to the flow,  a typical configuration adopted in 
order to investigate the problem of the LLBL formation at the Magnetopause. In this case, if the mean magnetic field angle 
remains relatively small we can assume that the ``{\em e}TF model'' \cite{ext_two-fluid_model} used 
here is still valid.  However, as already discussed, the  dynamics of the vortex chain generated by the K-H instability {advects}
the in-plane magnetic field producing regions where the magnetic field is locally amplified as for example nearby 
the vortex boundaries or in between two vortices undergoing pairing. Nevertheless, the region where the ``{\em e}TF model'' might fail, in 
the sense that the mean magnetic field is no longer ``almost perpendicular'' to the plane of the dynamics, is  mostly limited  to  
the vortex ribbons  while inside the vortices, where 
the anisotropy generation mechanism is the more efficient, the in-plane magnetic field remains small or even negligible. 
\begin{figure} [h!]
\includegraphics[width=3.7cm]{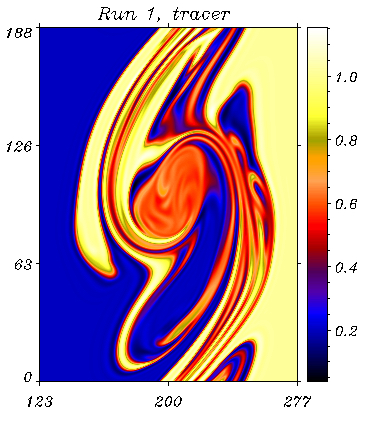}
\includegraphics[width=3.8cm]{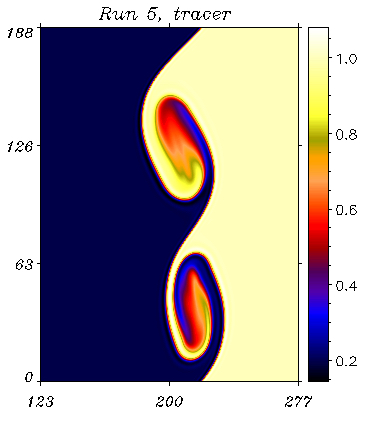}
\includegraphics[width=4.0cm]{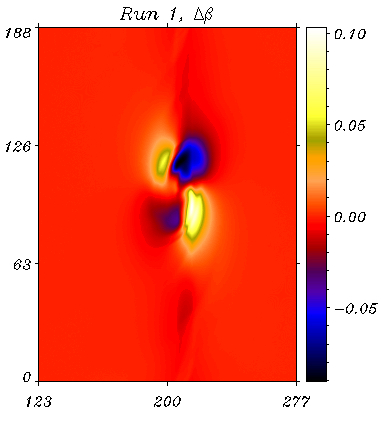}
\includegraphics[width=3.9cm]{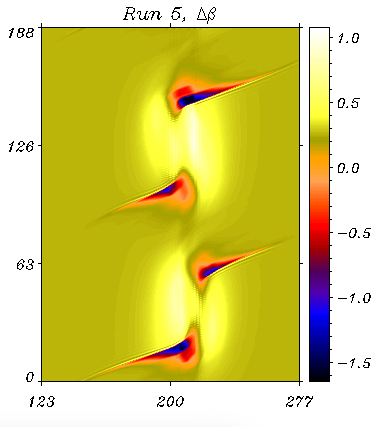}
 \caption{The strictly perpendicular magnetic field case. First two frames: the tracer, subsonic and supermagnetosonic  regime, Run 1 and 5, respectively. 
 Last two frames: the plasma anisotropy, subsonic and supermagnetosonic  regime, Run 1 and 5, respectively.}
 \label{fig7}
 \end{figure}

{ In order to prove  that the anisotropy generation is not an {artifact} of the model { being pushed towards  magnetic configurations
that are not non strictly perpendicular}, we have performed three runs, Run 1, 3 and 5,
 identical to the previous ones, but initializing 
the system with a strictly perpendicular magnetic field (see Table  \ref{tab:mach_uniform}). In Fig. \ref{fig7} we show 
the shaded iso-contours of the tracer, first two frames and of $\Delta \beta_{anis}$, last two frames for Run 1 and 5, 
respectively. The most important result is that in the supermagnetosonic  regime, run 5, the efficiency of the plasma in generating 
plasma anisotropy, see Fig. \ref{fig7} last frame, is {approximately} the same as in the case with an initial small in-plane magnetic 
field, run 6, discussed in Sec.\ref{magn_reg}. This is due to the fact that two robust vortices are generated by the KHI  
where plasma anisotropy is efficiently produced. A similar behavior is qualitatively and quantitatively observed { in} 
the intermediate regime, run 3 (for the sake of brevity not shown here). On the other hand, a significant difference is 
observed in the subsonic regime where the plasma  is now {much less prone to the onset of } secondary instabilities {and 
in particular of magnetic reconnection }(see Fig. \ref{fig7}, first frame). As a consequence, no small scale 
structures are generated { which are  the regions where in the non strictly perpendicular case, run 2, the anisotropy 
peaks are observed}.  

\section{Conclusions}\label{conclusions}

The generation of pressure anisotropy  during the nonlinear evolution of a low collisionality plasma  is being increasingly considered  as a key dynamical player in space plasma systems,  first of all in the heliosphere, but also in astrophysical systems as, e.g., (hot) accretion disks or galaxy clusters  as well as in laboratory systems e.g., in the context of magnetic reconnection}}.
In particular, distribution functions with different thermal pressure in the direction parallel or perpendicular to the magnetic field are today routinely detected in 
the solar wind. Moreover, structures emerging from the development of anisotropy instabilities  such as, e.g. mirror modes, are also often observed in the solar wind.
Recently, a strong theoretical and observational debate has occurred  about the role of the firehose and mirror instability (using by 
the parallel plasma $\beta$) in constraining the formation of solar wind proton temperature anisotropy. In fact, pressure anisotropy 
is an  important free energy source for relatively fast micro-instabilities that in turn can have a strong impact on the global structuring of the magnetic field or even on 
the shaping of the full system.

 A series of different mechanisms, involving energy transfer between  the parallel and the perpendicular direction, or preferential energy input from an external source 
 or energy loss e.g., due to cyclotron emission, can induce pressure anisotropy in a low collisionality plasma. 
Here  we have shown that the nonlinear dynamics of a vortex chain resulting from the development of the Kelvin Helmholtz instability driven by a sheared flow close to, or in, the supermagnetosonic  regime, can produce  a significant pressure anisotropy.    This result is obtained by integrating numerically, in a  two dimensional  plasma configuration, a  (two-fluid) dynamical model that includes a pressure tensor evolution equation explicitly.  The resulting  anisotropy is localized in the flow shear region. Pressure anisotropy, 
depending on its sign, can lead to the onset  either of the mirror or of the firehose instability.   
The physical mechanism that causes such an anisotropy  consists of two main steps: 

$i$)  first, the nonlinear evolution of the KH vortices compresses the plasma at  the vortex boundaries and between them  and  expands it  inside  the vortices, 

$ii$)  then,  the difference between  the   parallel and  the perpendicular polytropic  indices   in the pressure tensor equation  leads to  a different  variation  of the parallel and of the perpendicular pressure tensor components even  in the case of an  initially isotropic  pressure.

These  results  concerning the  spontaneous formation of pressure anisotropy  and  the onset of anisotropy instabilities, such as the  firehose or mirror instabilities, 
open up different possible scenarios in the nonlinear evolution of collisionless plasmas \cite{kunz2014}. In particular, not only the presence of pressure anisotropy can affect the KH dynamics  \cite{Prajapati2010}, but also an early development of secondary instabilities driven by pressure anisotropy may compete with the magnetofluid timescale of the vortex dynamics, 
potentially becoming the dominant process at the later stage of the nonlinear evolution. Many configurations where shear flow instabilities develop have 
been outlined in the literature. First of all, let us underline that direct signatures about the role of the KHI as a primary 
drive for secondary mechanisms leading to temperature anisotropy have been detected in the plasma sheet in the near-Earth magnetosphere by satellite observations \cite{NishinoAnnGeo2007} and that the relevance of shear flow configurations for anisotropy-driven instabilities has been outlined also at the heliopause \cite{KuznetsovAL1995}. Furthermore, signatures indicating  a possible development of a KHI have been observed in some merging galaxy groups \cite{RoedigerApJ2012} and at the sloshing cold fronts 
of some galaxy clusters~\cite{RoedigerApJ2013}. Velocity shear instabilities are indicated also to play a role between the hot intracluster gas and the moving 
galaxies where turbulent layers could be generated \cite{RoedigerMNRAS2013}. Finally, we note that anisotropy pressure development can play an important 
role also on the context of magnetic reconnection by, e.g. changing the symmetry of the system significantly and modifying the magnetic field configuration \cite{Cassak2015}. 
Therefore, the KHI and its associated secondary instabilities including pressure anisotropy instabilities, may provide a relevant mechanism for the dynamics and the generation of 
turbulent layers in many different environments, not limited to the LLBL magnetosphere case. 

We conclude by noticing that this analysis must be considered as a first step towards a full 3D kinetic treatment that has been adopted for t
he sake of mathematical and computational simplicity. Indeed, the present analysis is on the one hand well suited to identify the onset of such 
a mechanism of anisotropy generation but, on the other hand, it  cannot predict accurately its evolution once the firehose or mirror instability 
threshold is approaching, first of all because of the 2D configuration. We nevertheless underline that even within these limitations, our analysis 
is relevant to asses that shear flows in space and, more generally in collisionless magnetized plasmas, are an important energy source for 
the development of anisotropy instabilities at play during the further nonlinear dynamics.

\acknowledgments
The research leading to these results has received funding from the European Commissions Seventh Framework Programme (FP7/2007-2013) under 
the grant agreements 263340/SWIFF (www.swiff.eu). One of the authors  (FC) is glad to thank Dr. C. Cavazzoni (CINECA, Italy) for his essential contribution to code parallelisation and performance. We acknowledge the access to Supermuc machine at LRZ made available within the PRACE 
initiative receiving funding from the European Commissions Seventh Framework Programme (FP7/2007-2013) under Grant Agreement 
No. RI-283493, Project No. 2012071282. Part of the simulations presented in this work were carried out using the HYDRA supercomputer at 
the Rechenzentrum Garching (RZG), Germany.

\end{document}